\documentclass[prmaterials, reprint, aps, amsmath, amssymb]{revtex4-2}
\usepackage{amsmath}

\usepackage{siunitx}
\usepackage{lmodern} % Font size support
\usepackage[bookmarks, colorlinks=false]{hyperref}
\usepackage{mathtools}
\hypersetup{colorlinks=true, allcolors=blue}

\DeclareSIUnit\angstrom{\text {Å}}

\usepackage{bm}

\begin{document}

% VECTOR convention for bold upright symbol
\renewcommand{\vec}[1]{\mathbf{#1}}

% TENSOR convention for upright bold symbol
\renewcommand{\tensor}[1]{\vec{#1}}

% INTEGRAL element convention
\def\dint#1{\text{d}#1}
\def\grad{\text{grad}}

\title{Volume of a dislocation network}

\author{Max Boleininger}
	\email{max.boleininger@ukaea.uk}
	\affiliation{UK Atomic Energy Authority, Culham Centre for Fusion Energy, Oxfordshire OX14 3DB, United Kingdom}

\author{Sergei L. Dudarev}
	\email{sergei.dudarev@ukaea.uk}
	\affiliation{UK Atomic Energy Authority, Culham Centre for Fusion Energy, Oxfordshire OX14 3DB, United Kingdom}

\author{Daniel R. Mason}
	\email{daniel.mason@ukaea.uk}
	\affiliation{UK Atomic Energy Authority, Culham Centre for Fusion Energy, Oxfordshire OX14 3DB, United Kingdom}

\author{Enrique Mart\'{i}nez}
	\email{enrique@clemson.edu}
	\affiliation{Department of Materials Science and Engineering, Clemson University, Sirrine Hall, Clemson, SC 29634, USA}
	
\date{\today}
\begin{abstract}
We derive a simple analytical line integral expression for the relaxation volume tensor of an arbitrary interconnected dislocation network. This quantity determines the magnitude of dislocation contribution to the dimensional changes and volumetric swelling of a material, and highlights the fundamental dual role of dislocations as sources of internal strain as well as carriers of plastic deformation. To illustrate applications of the method, we compute the relaxation volume of a stacking fault tetrahedron, a defect commonly occurring in fcc metals; the volume of an unusual tetrahedral configuration formed by the $(a/2)\langle 111\rangle$ and $a\langle 001\rangle$ dislocations in a bcc metal; and estimate the relative contribution of extended dislocations to the volume relaxation of heavily irradiated tungsten.  
\end{abstract}
\maketitle

%\begin{keywords}
%dislocation network \sep relaxation volume \sep volume swelling \sep linear elasticity theory \sep radiation damage
%\end{keywords}

\maketitle

\section{Introduction}

%lets not get into these complications for the moment, but we will have to sort this out at some point \textcolor{blue}{It would be appropriate to say that the question about the volume of network is fairly particular. For example, there is a quantity like geometrically necessary dislocations, to quantify bending of crystal lattice observed in EBSD. Volume of dislocation network question refers to non-bending type of lattice deformation.Unclear whether depending on the type of dislocation loop structure we are going to arrive at a non-isotropic volumetric expansion. SLD: the problem of GND is a little artificial since a spatially heterogeneous distribution of point defects would also produce local bending, there might be a problem of uniqueness here } 

A material exposed to neutron or ion irradiation changes its volume \cite{Cawthorne1967, Garner2000, Garner2020, Bhattacharya2020} and, in some cases, its anisotropic spatial dimensions \cite{Holt1988}. If the exposure of the material to a flux of energetic particles is low and the defects produced by irradiation can be treated as a dilute gas of localised centres of lattice distortion \cite{Leibfried1978}, the degree of volumetric expansion or contraction \cite{Simmons1958, Simmons1960, Cawthorne1967, Hertz1973} can be estimated from the relaxation volumes of individual defects \cite{dudarev2018multi, dudarev2018elastic, mason2019relaxation, reali2021macroscopic}. These volumes can now be accurately computed using {\it ab initio} methods \cite{Domain2005, Varvenne2013, Clouet2018, MaPRM2019a, ma2021nonuniversal}. 

However, a conceptual difficulty arises if the exposure to neutron or ion irradiation exceeds approximately 0.1 dpa \cite{Derlet2020,Mason2020}. In this high exposure limit, defects form complex dense microstructures \cite{Hernandez2008}, with vacancies at elevated temperatures coalescing into mesoscopic voids \cite{Cawthorne1967,Garner2000,Garner2020,Bhattacharya2020}. The self-interstitial defects cluster into dislocation loops, which then form rafts \cite{Jenkins2009,Dudarev2010,li2019diffusion}, agglomerate into larger loops \cite{Jenkins2009,Arakawa2011}, and eventually evolve into an interconnected network of dislocation lines \cite{Derlet2020,Mason2020}. 

The relaxation volume of a macroscopic void can be evaluated from its surface stress or surface free energy \cite{mason2019relaxation}.  In the absence of internal pressure from helium or hydrogen gas filling the void, the relaxation volume of a void is negative.  As a result, vacancies and voids distributed in a material produce negative lattice strain, readily observed using X-ray diffraction \cite{Simmons1960, Hertz1973, mason2019relaxation, Mason2020}. At the same time, it is well known that materials swell, i.e. increase their macroscopic volume, as a result of exposure to irradiation \cite{Cawthorne1967, Garner2000, Garner2020}. In the limit of low radiation exposure, this can be readily explained by the fact the relaxation volumes of self-interstitial defects, produced simultaneously with vacancies during irradiation, are positive and significantly larger than the (negative) relaxation volumes of vacancies \cite{MaPRM2019a,ma2021nonuniversal}. But in the limit of relatively high exposure to irradiation this argument no longer applies, since the majority of self-interstitial atom defects now agglomerate into large dislocation loops and an extended network of interconnected dislocations \cite{Derlet2020,Mason2020}, effectively making them a part of the regular crystal lattice, and hence barely detectable by electron microscopy or X-ray diffuse scattering. In fcc metals, vacancies form stacking fault tetrahedra \cite{Singh1993, Kiritani2000, Marian2009, Martinez2015}, which are dislocation configurations.

Extensive dislocation networks also form during severe plastic deformation of materials, especially during rapid deformation \cite{Kiritani2003a,Kiritani2003b,Marian2004}, where the atomic processes resemble those occurring in high-energy collision cascades initiated by the incident high-energy ions or neutrons \cite{Calder2010,Sand2013}. 

To relate the complexity of dislocation structures and dislocation networks to the macroscopic dimensional changes that they produce, and enable including dislocations in the finite-element model formalism for the numerical treatment of dimensional changes and radiation swelling of nuclear fusion reactor components \cite{dudarev2018multi, reali2021macroscopic}, we propose a method for computing relaxation volume tensors and relaxation volumes of arbitrary dislocation configurations, and show how to evaluate the degree of volumetric expansion in a material containing an interconnected network of dislocation lines.

\section{Relaxation volume of an isolated defect or a dislocation loop}\label{sec:relvol}

We start by revisiting the concept of relaxation volume in linear elasticity theory, casting it into a framework coherent with the established eigenstrain \cite{mura2013micromechanics} and dipole tensor \cite{Leibfried1978} formalisms, and treating it from the perspective of recent multi-scale applications \cite{dudarev2018multi, reali2021macroscopic}.

Consider a crystal occupying a finite volume in three-dimensional space, subject to the traction-free surface boundary conditions. The relaxation volume is defined as the change of the volume enclosed by the external surface of the crystal, caused by the strain inside it \cite{Leibfried1978,Puchala2008}. Hence, the relaxation volume tensor $\tensor\Omega$ is given by the surface integral of the displacement field $u_i(\vec x)$ over the volume boundary $S$ with normal vector $n_j(\vec x)$, namely
\begin{equation}\label{eq:relaxationtensordefinition}
    \Omega_{ij} = \frac{1}{2} \int_{S} \left[u_i ({\vec x})n_j ({\vec x}) + u_j ({\vec x}) n_i({\vec x})\right] \dint{S}.
\end{equation}
This quantity is a tensorial measure of the volume change to first order in displacements \cite{reali2021macroscopic}, with the relaxation volume $\Omega$ given by the trace
\begin{equation}
    \Omega = {\rm Tr}(\tensor\Omega) = \Omega_{ii},
\end{equation}
where summation over repeated indices is implied, $\Omega_{ii}=\Omega_{11}+\Omega_{22}+\Omega_{33}$. Note that only the diagonal elements of $\Omega_{ij}$ contribute to the relaxation volume, as shear distortions conserve the volume of the material.

Applying the divergence theorem to \eqref{eq:relaxationtensordefinition}, we arrive at \cite{Leibfried1978,Puchala2008}
\begin{equation}\label{eq:relaxationtensor}
    \Omega_{ij} = \int_{V} \varepsilon_{ij}({\bf x})\, \dint{^3x},
\end{equation}
where $\varepsilon_{ij} = \frac{1}{2} (u_{i,j} + u_{j,i})$ is the strain tensor, with subscripts after a comma denoting differentiation ($f_{,l} = \partial f/\partial x_l$). The notion of the relaxation volume tensor enables quantifying anisotropic dimensional changes, as its eigenvalues represent variations of volume in the principal directions of the tensor.

We are interested in the case where the elastic strain is originating from the defects inside the crystal. In Mura's formalism \cite{mura2013micromechanics}, the defects are described by a spatially-varying field of eigenstrain $\varepsilon_{ij}^*({\bf x})$, which acts as a source of elastic deformation. The strain tensor is thereby given by the sum of two terms
\begin{equation}\label{eq:muraeigen}
    \varepsilon_{ij}(\vec{x}) = \varepsilon_{ij}^\mathrm{el.}(\vec{x}) + \varepsilon_{ij}^*(\vec{x}),
\end{equation}
where the elastic strain $\varepsilon_{ij}^\mathrm{el.}({\bf x})$ also includes the effect of boundary conditions. In the particular case where the surfaces are free of tractions, the elastic strain tensor averages to zero over the volume of the material \cite{Indenbom1992,el2000boundary}. Consequently, the relaxation volume tensor is given by the volume integral of the eigenstrain \cite{reali2021macroscopic}: 
\begin{equation}\label{eq:omegaeigen}
    \Omega_{ij} = \int_{V} \varepsilon_{ij}^*(\vec{x})\, \dint{^3 x}
\end{equation}

In applications, it is necessary to express the eigenstrain in terms of the specific content of defects in the crystal. For a point $\vec{x}$ inside the crystal, we can evaluate the eigenstrain from the body force $f^*_i$ by using the elastostatic equilibrium condition $\sigma_{ij,j} = -f_i$, {\it viz.}
\begin{equation}\label{eq:bodyforce}
    C_{ijkl} \varepsilon^*_{kl,j}(\vec{x}) = -f^*_i(\vec{x}),
\end{equation}
where $C_{ijkl}$ is the stiffness tensor. The displacement field originating from the body force is given by the volume integral
\begin{equation}\label{eq:greensdisplacementfull}
\begin{aligned}
u_{i}(\vec{x}) 
    &= \phantom{-} \int G_{ij}(\vec{x}-\vec{x}') f^*_j (\vec{x}')\, \dint{^3x'} \\
    &= - \int G_{ij,k}(\vec{x}-\vec{x}') C_{jklm} \varepsilon^*_{lm}(\vec{x}')\, \dint{^3x'},
\end{aligned}
\end{equation}
where $G_{ij}(\vec{x}-\vec{x}')$ is the elastic Green's function of the infinite body. Note that the above formula is not a solution to the general boundary-value problem, as boundary conditions are not accounted for, see Refs.~\cite{eshelby1954distortion, kossevich1999crystal} for detail.

If the eigenstrain is associated with a defect localised in a region near point $\vec{R}$ in the crystal, the asymptotic form of the displacement field is found by expanding the elastic Green function to lowest order in $\vec{x}'$ at $\vec{R}$,
\begin{equation}\label{eq:greensdisplacementrelvol}
    u_{i}({\vec x}) = -C_{jklm} \Omega_{lm} G_{ij,k}(\vec{x}-\vec{R}),
\end{equation}
where $\Omega_{lm}$ is the relaxation volume tensor given by Eq.~\eqref{eq:omegaeigen}. This gives rise to the expression known from the dipole tensor formalism \cite{Leibfried1978, dudarev2018elastic}, namely
\begin{equation}\label{eq:greensdisplacement}
    u_{i}(\vec{x}) = - P_{jk} G_{ij,k}(\vec{x} - \vec {R}),
\end{equation}
where the relaxation volume and elastic dipole tensor $P_{ij}$ of the defect are related through
\begin{eqnarray}
    \Omega_{ij} &=& S_{ijkl} P_{kl} \label{eq:ptoomega} \\
    P_{ij}      &=& C_{ijkl} \Omega_{kl} \label{eq:omegatop},
\end{eqnarray}
and $S_{ijkl}$ is the compliance tensor \footnote{The compliance tensor is defined by its relation to the stiffness tensor, namely $C_{ijkl} S_{klmn} = \frac{1}{2}( \delta_{im}\delta_{jn} + \delta_{in}\delta_{jm}$).}. Owing to the symmetry properties of the stiffness and compliance tensors \cite{sutton2020physics}, both the relaxation volume and elastic dipole tensors are symmetric.
%Equation \eqref{eq:greensdisplacement} is the first term in the multipole expansion, dominating the elastic field of a defect at large distances from it \cite{Leibfried1978}. 
The notion of elastic dipole tensor enables evaluating the energy of interaction between the defect and an external slowly spatially varying strain field $\varepsilon_{ij}'({\vec x})$ to first order in the size of the defect as $E({\vec x}) = -P_{ij}\varepsilon_{ij}'({\vec x})$, see \cite{Clouet2008,MaPRM2021finiteT}. For a detailed analysis of how elastic dipole tensors, formation \footnote{Formation volumes of defects, explored for example in Ref. \cite{MaPRM2019b}, differ from relaxation volumes in that they do not solely refer to effects of lattice distortions. A calculation of the formation volume of a defect is based on a different definition, resulting in that the numerical values of formation volumes of a vacancy and a self-interstitial atom defect are similar.} and relaxation volumes of small isolated defects can be evaluated from atomistic simulations, we refer an interested reader to Refs.~\cite{Varvenne2013,Varvenne2017,dudarev2018elastic,mason2019relaxation,MaPRM2019a,MaPRM2019b,ma2021nonuniversal,Clouet2018}. 
%
%\begin{eqnarray}
%    \Omega_{ij} &=& \Omega _{ji} \label{eq:Omega_symm} \\
%    P_{ij}      &=& P_{ji}. \label{eq:P_symm}
%\end{eqnarray}
%

To relate the relaxation volume tensor of a localised defect to Mura's eigenstrain, we define the density of relaxation volumes of defects \cite{dudarev2018multi} similarly to the notion of charge density in electrodynamics, see e.g. Eq.~(28.1) of Ref.~\cite{LandauTheoryofFields}, 
\begin{equation}\label{eq:volume_density}
    \omega _{kl}({\bf x})=\sum _a \Omega ^{(a)}_{kl}\delta ({\bf x}-{\vec R}_a),
\end{equation}
where summation is performed over the positions of individual defects ${\vec R}_a$, and $\delta(\vec{x})$ is the Dirac delta function defined as $\delta ({\bf x})=0$ for ${\bf x}\ne 0$ and $\int \delta ({\bf x}) d^3x=1$. The relaxation volume tensor \eqref{eq:omegaeigen} can now be expressed as an integral of $\omega_{kl}({\bf x})$ over the volume of the material $\Omega _{kl}=\int_V \omega _{kl}({\bf x})\, \dint {^3x}$. Using \eqref{eq:greensdisplacementrelvol} and \eqref{eq:volume_density}, the field of displacements generated by a spatial distribution of defects can be evaluated as an integral of the density of relaxation volumes \cite{mura2013micromechanics,reali2021macroscopic}
\begin{equation}\label{eq:distributed_displacement}
u_{i}(\vec{x}) = - \int G_{ij,k}(\vec{x}-\vec{x}') C_{jklm} \omega_{lm}(\vec{x}')\, \dint{^3x'}.
\end{equation}
By comparing the above with Eq.~\eqref{eq:greensdisplacementfull}, we arrive at the \textit{defect eigenstrain theorem} \cite{reali2021macroscopic}, which is the realisation that the density of relaxation volume tensors is equivalent to the eigenstrain,
\begin{equation}\label{eq:eigenstrain-theorem}
    \varepsilon^*_{ij}(\vec{x}) \equiv \omega_{ij}(\vec{x}).
\end{equation}
Provided that the spatial distribution of crystal defects and their relaxation volume tensors are known, the resulting macroscopic stresses can be evaluated using conventional linear elasticity theory \cite{dudarev2018multi, reali2021macroscopic}. This theorem highlights the intrinsic multi-scale character of elastic forces originating from crystal defects. 

While the relaxation volume and dipole tensors of point defects and small defect clusters can be readily obtained from atomistic or density functional theory simulations \cite{MaPRM2019a,MaPRM2019b,ma2021nonuniversal}, it is not immediately clear how to identify the relaxation volume of dislocation-type defects, in particular when the dislocations form a complex interconnected network. In what follows, we treat the problem in the linear elasticity approximation and begin by exploring the relaxation volume produced by an isolated dislocation loop, a prototypical building block of a general dislocation network. 

A dislocation curve is defined as the boundary of a cut surface inside the crystal bulk, where the sides of the cut are displaced by the Burgers vector $\vec{b}$ with respect to each another. As cut surfaces are generally open surfaces, dislocation lines cannot have loose ends inside the crystal. Not accounting for closures along free surfaces or grain boundaries, dislocation lines must terminate either at dislocation junctions or by looping back onto themselves. In the following, we refer to a dislocation network containing no loose line segments as \textit{closed}. Following the above definition, the density of relaxation volumes or, equivalently, the eigenstrain of a dislocation loop, is given by \cite{Po2018}
\begin{equation}\label{eq:volume_density_loop}
\omega_{ij}(\vec{x}) = \frac{1}{2} \int_S \left(
    b_i \dint{S}_j + b_j \dint{S}_i\right) \delta\left[\vec{x}-\vec{r}_S(v,w)\right],
\end{equation}
where
\begin{equation*}
\dint S_n =\left( \frac{\partial {\bf r}_S}{\partial v} \times  
                  \frac{\partial {\bf r}_S}{\partial w}\right)_n \dint v \dint w
\end{equation*}
defines the bounding surface of a dislocation loop in a parameterised form through the vector function ${\bf r}_S(v,w)$  \cite{KornKorn1968}. In the linear elasticity approximation, the choice of a particular shape of the bounding surface ${\bf r}_S(v,w)$ is immaterial, since strain and stress are unique and continuous functions defined solely by the position of the dislocation line at the perimeter of the bounding surface of the loop \cite{landau1986theory}. However, if one treats the underlying discrete atomic crystal structure of the material, the variations of the shape of the bounding surface can be detected and analysed, for example by identifying the extrema of strain associated with the individual crowdion defects constituting a dislocation loop \cite{boleininger2019continuum, boleininger2020ultraviolet}. The fact that a perfect dislocation loop is an assembly of crowdion point defects explains why, if we neglect the core effects, the volume of a loop equals the number of defects forming it, multiplied by the volume of an atom in the crystal lattice. We refer an interested reader to Ref.~\cite{mason2019relaxation} for a detailed analysis of scaling relations for the relaxation volumes of larger defects, derived from atomistic simulations.

Evaluating the volume integral \eqref{eq:omegaeigen} for the eigenstrain \eqref{eq:volume_density_loop}, we arrive at the relaxation volume tensor of an isolated dislocation loop, see also equations (27.11) and (27.12) by Landau and Lifshitz \cite{landau1986theory},
\begin{equation}\label{eq:relaxationtensordisloc}
    \Omega_{ij} = \frac{1}{2} \left( b_i A_j + b_j A_i \right),
\end{equation}
where $A_i$ is the {\it vector} area of the loop. The vector area is defined either as an integral over the bounding surface of the cut or, through the Stokes theorem, as a line integral over the boundary $\Gamma$ of the cut \cite{landau1986theory},
\begin{equation}\label{eq:area_vector}
    A_i = \int _{S} n_i({\bf x})\, \dint{S} = \frac{1}{2} \oint_{\Gamma} \epsilon_{ikl} r_k\, \dint{r_l},
\end{equation}
where $n_i({\bf x})$ is the surface unit normal vector, $\epsilon_{ikl}$ is the Levi-Civita tensor, and ${\bf r}=\{r_k\}$ is a coordinate of a point on a dislocation line forming the perimeter of the dislocation loop. In vector notation, equation \eqref{eq:area_vector} is \cite{Rovelli2018}
\begin{equation} \label{eq:area_vector1}
      {\bf A} = \frac{1}{2} \oint_{\Gamma} {\bf r} \times  \dint{{\bf r}}.
\end{equation}
This definition is invariant with respect to the choice of the origin of the Cartesian system of coordinates, since a translation of the origin by a constant vector ${\bf r}_0$ adds to \eqref{eq:area_vector1} a term 
\begin{equation*}
\frac{1}{2} \oint_{\Gamma} {\bf r}_0 \times \dint{{\bf r}}=\frac{1}{2} {\bf r}_0\times \left(\oint_{\Gamma} \dint{{\bf r}}\right),
\end{equation*}
which vanishes for a closed loop since ${\oint_{\Gamma} \dint {\bf r}}=0$. 

The relaxation volume of a dislocation loop is given by the trace of \eqref{eq:relaxationtensordisloc} and hence equals the scalar product of the Burgers vector of the loop and its vector area, cf. Eq.~(43) of Ref.~\cite{Dederichs1973} and Eq.~(12) of Ref.~\cite{dudarev2018elastic}
\begin{equation}
\Omega  =  {\bf{b}} \cdot {\bf{A}}
        = \frac{1}{2} \oint_{\Gamma} {\bf b}\cdot ({\bf r} \times  \dint{{\bf r}})=\frac{1}{2} \oint_{\Gamma} ({\bf b}\times {\bf r}) \cdot \dint{{\bf r}}.\label{eq:volume_loop}
\end{equation}
The relaxation volume of an interstitial loop is positive $({\bf b}\cdot{\bf A})>0$, whereas the relaxation volume of a vacancy loop is negative $({\bf b}\cdot{\bf A})<0$. The relaxation volume of a loop formed by a pure shear-type deformation is equal to zero
$({\bf b}\cdot{\bf A})=0$.

The elastic dipole tensor of a dislocation loop can be readily found from the relaxation volume tensor using equation \eqref{eq:omegatop} as
\begin{equation}
    P_{ij} = C_{ijkl} A_k b_l.\label{eq:dipole_tensor_loop}
\end{equation}
In the above equation, there is no need to symmetrise the expression in the right-hand side with respect to $i$ and $j$ since the symmetry properties of $P_{ij}$ automatically follow from the symmetry of $C_{ijkl}$. Similarly, there is no need to use the symmetrised form (\ref{eq:relaxationtensordisloc}) in the right-hand side of the equation. 

As a slight digression, we note that the Burgers formula for the displacement field of a curved dislocation line can be obtained by substituting \eqref{eq:volume_density_loop} into \eqref{eq:distributed_displacement}
\begin{equation}\label{eq:Burgers_formula}
    u_i({\bf x})=-C_{klmn}b_m \int _{S}G_{ik,l}({\bf x}-{\bf x}')\, \dint{S'_n}, 
\end{equation}
which is equivalent to Eq.~(9.3) by Teodosiu \cite{Teodosiu1977}, Eq.~(4-6) by Anderson, Hirth and Lothe \cite{anderson2017theory}, and Eq.~(27.10) by Landau and Lifshitz \cite{landau1986theory}. In \eqref{eq:Burgers_formula}, integration is performed over the bounding surface of a dislocation loop. The differentiation of the elastic Green's function in \eqref{eq:Burgers_formula} is performed with respect to coordinate ${\bf x}$, namely
$G_{ik,l}({\bf x}-{\bf x}')={\partial}G_{ik}({\bf x}-{\bf x}')/\partial x_l$, and the sign convention in the definition of the Burgers vector in equation (\ref{eq:Burgers_formula}) above is consistent with Refs.~\cite{landau1986theory,Teodosiu1977,Dederichs1973}. If one adopts the alternative definition of the Burgers vector used in Ref.~\cite{anderson2017theory}, the sign before equation \eqref{eq:Burgers_formula} must be amended from minus to plus. The observed physical quantities, for example relaxation volumes, are independent of the Burgers vector sign convention, and the volume of a self-interstitial dislocation loop is always positive whereas the volume of a vacancy loop is negative.  

A remarkable aspect of equations \eqref{eq:volume_density_loop} and \eqref{eq:volume_loop} is that, on the one hand, they show that the fundamental origin of the positive (or negative) volume of a dislocation loop is associated with the extra (or missing) material at the bounding surface $S$ of the loop, illustrated by equation \eqref{eq:volume_density_loop} and readily identified in atomistic models \cite{boleininger2018msfk,dudarev2003coherent}.  On the other hand, equation \eqref{eq:volume_loop} shows that the problem of evaluation of the volume of a dislocation loop can be conveniently reduced to computing a contour integral along the dislocation line at the perimeter of a dislocation loop. This result, replacing the procedure of counting extra or missing atoms over the surface of a dislocation loop by a line integration over its perimeter, enables generalising the formalism developed above for an isolated loop to an arbitrary configuration of entangled dislocations.  

Equations \eqref{eq:relaxationtensordisloc}, \eqref{eq:volume_loop} and \eqref{eq:dipole_tensor_loop} show how to evaluate the relaxation volume and elastic dipole tensors of an individual dislocation loop. The case of a dislocation network, featuring complex configurations of dislocation lines and dislocation junctions, requires a special consideration. The formalism for evaluating the relaxation volume tensor of a dislocation network is given in the next section of the paper.  

\section{Relaxation volume of a dislocation network}\label{sec:relnetwork}

\begin{figure*}[t]
\raggedright
\includegraphics[width=\textwidth]{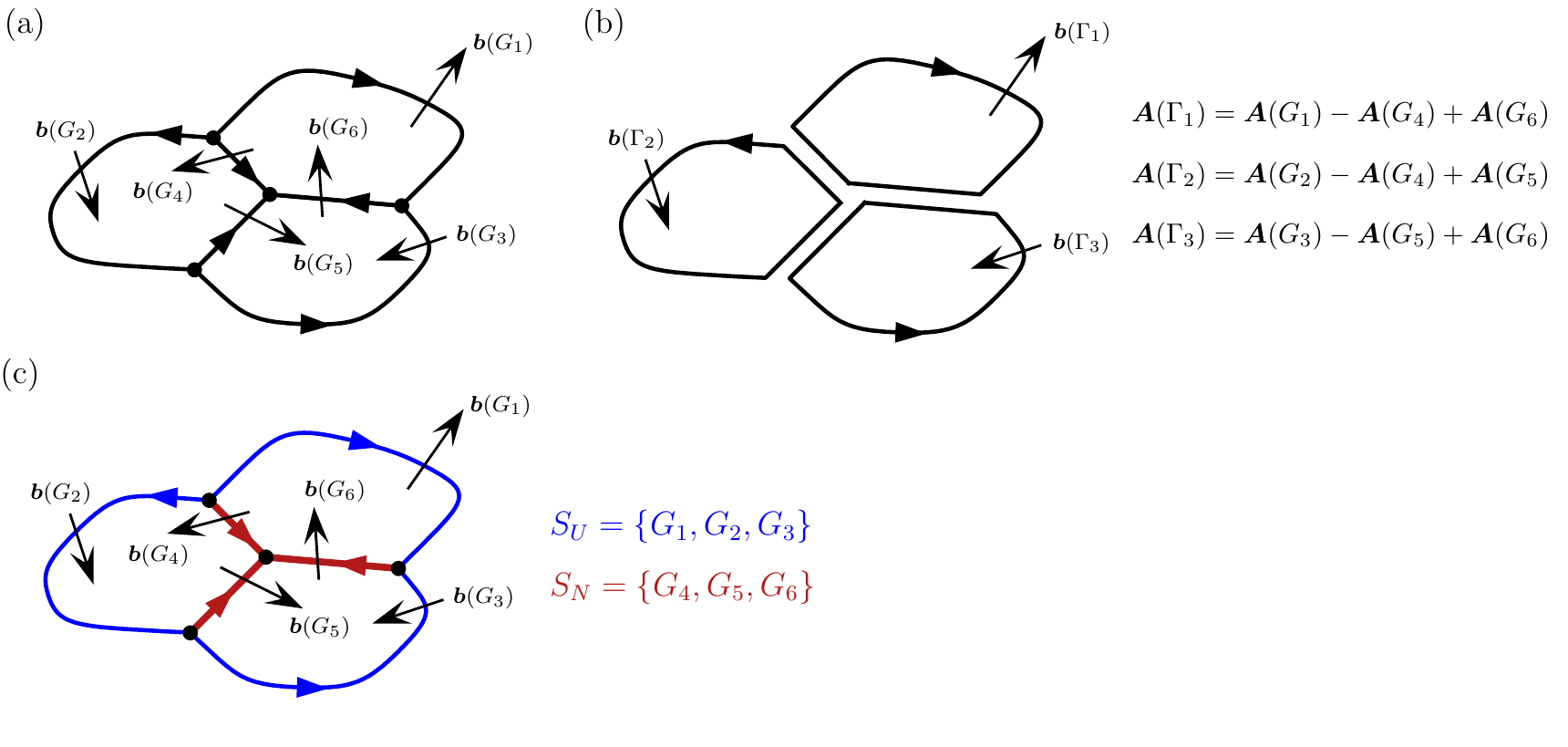}
\caption{
(a) Example of a closed dislocation network consisting of line segments $G_i$ with Burgers vectors $\vec{b}(G_i)$ (arrows). Triangles indicate line sense. %
(b) The same network represented as a superposition of closed loops $\Gamma_i$. Also given is the relation between area vectors $A(\Gamma_i)$ and $A(G_i)$. %
(c) Closed loops are split into their constitutive line segments, which are then grouped according to whether they appear uniquely or not, into sets $S_U$ and $S_N$, respectively. The decomposition is used to prove that all three representations yield the same dipole tensor.
\label{fig:decomposition}
}
\end{figure*}

A dislocation network is defined by a set of dislocation junctions connected by dislocation line segments $G_n$ with Burgers vectors ${\bf b}(G_n)\equiv b_i(G_n)$. A closed dislocation network can be equivalently represented as a superposition of closed dislocation contours $\Gamma_\alpha$ with Burgers vectors $b_i(\Gamma_\alpha)$ and area vectors $A_i{(\Gamma_\alpha)}$ by splitting the dislocation segments according to the conservation rule of Burgers vector at junctions, analogous to the Kirchhoff law of electric current flow \cite{anderson2017theory}. The dipole tensor of the network in the linear elasticity approximation is therefore given by the sum of dipole tensors of the closed dislocation contours, namely
\begin{equation}
    P_{ij} = C_{ijkl} \sum_{\Gamma_\alpha} b_k{(\Gamma_\alpha)} A_l{(\Gamma_\alpha)}.
\end{equation}
%
%Note that $P_{ij}=P_{ji}$ due to the symmetry property of the stiffness tensor  $C_{ijkl}=C_{jikl}$ \cite{sutton2020physics}.
While the above expression shows that the dipole tensor of a closed dislocation network is unambiguously defined, it is in practice cumbersome to partition the network into individual contours. We shall instead reformulate the problem in terms of dislocation line segments by reversing the argument. 

First, similarly to \eqref{eq:area_vector} and \eqref{eq:area_vector1}, we split the area vectors of closed loops $\Gamma_\alpha$ into area vectors of their constituent line segments $G_n$,
\begin{equation}\label{eq:areadecompose}
    A_i{(\Gamma_\alpha)} = \sum_{G_n \in \Gamma_\alpha} A_i^\alpha{(G_n)},
\end{equation}
where $A_i^\alpha{(G_n)} = \pm A_i{(G_n)}$ as the line sense of segment $G_n$ may not match that of the loop $\Gamma_\alpha$, leading to the dipole tensor
\begin{equation}\label{eq:decompose1}
    P_{ij} = C_{ijkl} \sum_{\Gamma_\alpha} \sum_{G_n \in \Gamma_\alpha} 
            b_k{(\Gamma_\alpha)} A_l^\alpha{(G_n)}.
\end{equation}
This decomposition is exact as integration is a linear operation, with the caveat that the $A_i{(G_n)}$ line integrals are not closed anymore. 

Some new line segments were introduced in order to form the closed loop representation, which are now being integrated over multiple times. Defining $S_U$ as the set of line segments being integrated over once and $S_N$ as the set of line segments being integrated over several times, Eq.~\eqref{eq:decompose1} is reordered as
\begin{equation}
\begin{aligned}
    P_{ij}  = \phantom{+} &C_{ijkl} \sum_{G_n \in S_U} 
                                b_k{(G_n)} A_l{(G_n)} \\
                        + &C_{ijkl} \sum_{G_n \in S_N} \sum_{\Gamma_\alpha \in M(G_n)} 
                                b_k{(\Gamma_\alpha)}  A_l^\alpha{(G_n)},
\end{aligned}
\end{equation}
where summation over the set 
\begin{equation}
M(G_n) = \{ \Gamma_\alpha\, |\, G_n \in \Gamma_\alpha\}
\end{equation}
counts all appearances of segment $G_n \in S_N$. We refer to Fig.~\ref{fig:decomposition} for a visual example of the decomposition. 

Next, recalling that due to the conservation of the Burgers vector at junctions
\begin{equation}
\sum_{\Gamma_\alpha \in M(G_n)} b_k{(\Gamma_\alpha)}  A_l^\alpha({G_n}) = b_k{(G_n)} A_l({G_n}),
\end{equation}
as well as that the union of unique and non-uniquely appearing segments contains all the line segments of the network exactly once, we arrive at the central result:
\begin{equation}\label{eq:dipoletcurve}
    P_{ij} = C_{ijkl} \sum_{G_n} b_k{(G_n)} A_l{(G_n)}.
\end{equation}
Comparing Eq.~\eqref{eq:dipoletcurve} with Eq.~\eqref{eq:ptoomega}, we find the relaxation volume tensor
\begin{equation}\label{eq:relaxtcurve}
    \Omega_{ij} = \frac{1}{2} \sum_{G_n} \big[
        b_i{(G_n)} A_j{(G_n)} + b_j{(G_n)} A_i{(G_n)} \big],
\end{equation}
the trace of which yields the relaxation volume of the network  
\begin{equation}\label{eq:relaxcurve}
    \Omega = \sum_{G_n} b_i{(G_n)} A_i{(G_n)}.
\end{equation}
Since the evaluation of areas $A_i{(G_n)}$ can be reduced to the line integrals along the dislocations, the dipole tensor, the relaxation volume tensor, and the relaxation volume of a closed dislocation network can all be obtained by the integration over the individual dislocation line segments. 

In computational dislocation dynamics codes and post-processing algorithms for detecting dislocations, dislocation lines are commonly represented by a set of piecewise {\it directionally-ordered} linear segments \cite{dupuy2002study}, satisfying Kirchhoff's Burgers vector conservation law at dislocation junctions. Each linear segment $C_n$ is defined by its Burgers vector $b_i^{(n)}$, the starting point $p_i^{(n)}$ and the end point $q_i^{(n)}$. Noting that an element of the vector area associated with an individual straight segment equals
\begin{equation*}
{\bf A}^{(n)} = \frac{1}{4} \left({\bf q}^{(n)}+{\bf p}^{(n)}\right)
    \times \left({\bf q}^{(n)}-{\bf p}^{(n)}\right)
    = \frac{1}{2}{\bf p}^{(n)}\times {\bf q}^{(n)},
\end{equation*}
we can simplify the expression for the dipole tensor of a dislocation network \eqref{eq:dipoletcurve} as
\begin{equation}\label{eq:dipoleline}
    P_{ij} = \frac{1}{2} C_{ijkl} \epsilon_{luv} \sum_{n} b_k^{(n)} p_{u}^{(n)} q_{v}^{(n)}.
\end{equation}
The relaxation volume tensor of a dislocation network described by straight dislocation line segments is now 
\begin{equation}\label{eq:reltensorline}
    \Omega_{ij} = \frac{1}{4} \sum_{n} \left[
        b_i^{(n)} \epsilon_{juv} + b_j^{(n)} \epsilon_{iuv}
    \right] p_{u}^{(n)} q_{v}^{(n)},
\end{equation}
with the corresponding total relaxation volume given by  
\begin{equation}\label{eq:relaxline}
\begin{aligned}
    \Omega  &= \frac{1}{2}\epsilon_{iuv} \sum_{n} b_i^{(n)} p_{u}^{(n)} q_{v}^{(n)} \\
            &= \frac{1}{2}\sum_{n} {\bf b}^{(n)} \cdot ({\bf p}^{(n)}\times {\bf q}^{(n)}),
\end{aligned}
\end{equation}
where ${\bf p}^{(n)}$ and ${\bf q}^{(n)}$ are the coordinates of the start and end points of straight dislocation segments.
For a set of arbitrary curved dislocation segments used in dislocation dynamics simulations \cite{Ghoniem1999,Ghoniem2000,dupuy2002study,Po2014,Po2018}, the relaxation volume can be written as a sum of line integrals along the dislocation segments, linking dislocation junctions
\begin{equation}\label{eq:volume_network_integral}
\begin{aligned}
\Omega  &= \frac{1}{2}\epsilon _{iuv} \sum_{n} b_i^{(n)} \int _{(n)}r_u  \dint {r_v} \\
        &= \frac{1}{2} \sum_{n} \int_{(n)} ({\bf b}^{(n)} \times {\bf r}) \cdot {\dint {\bf r}} \\
        &= \frac{1}{2} \sum_{n} {\bf b}^{(n)} \cdot \int _{(n)} ( {\bf r} \times \dint {\bf r} ),
\end{aligned}
\end{equation}
where the direction of line integration along each curved segment follows the direction of the corresponding dislocation, consistent with the law of conservation of the Burgers vector at the dislocation junctions. Equation \eqref{eq:volume_network_integral} is one of the central results of our study, and for an individual dislocation loop it reduces to \eqref{eq:volume_loop}.

To show that expression \eqref{eq:volume_network_integral} is invariant with respect to an arbitrary translation of the Cartesian system of coordinates, we note that such a translation amounts to adding an extra term
\begin{equation}\label{eq:proof_invariance}
\begin{aligned}
   &\frac{1}{2}\sum_{n} {\bf b}^{(n)} \cdot 
            \left({\bf r}_0\times \int _{(n)} \dint{\bf r} \right) \\
 = &\frac{1}{2}\sum_{s} {\bf b}^{(s)} \cdot 
            \left({\bf r}_0\times \left[{\bf r}^{(s)}_2-{\bf r}^{(s)}_1\right] \right),
\end{aligned}
\end{equation}
where ${\bf r}_0$ is the translation vector, and summation over $s$ is performed over the full arbitrarily curved segments of the dislocation network linking the dislocation junctions, with ${\bf r}^{(s)}_2$ and ${\bf r}^{(s)}_1$ being the coordinates of the two junctions at the start and the end of a full segment $s$. Since the choice of the start and end points of a full segment reflects the sense of direction of integration along the segment, each junction enters the sum in \eqref{eq:proof_invariance} with a plus or minus sign. This sign rule, depending on whether a segment enters or leaves the junction, in combination with the Burgers vector conservation law at the junctions, ensures that the various terms cancel each other and the sum in \eqref{eq:proof_invariance} vanishes, confirming the translational invariance of expression \eqref{eq:volume_network_integral}.

\section{Periodic boundary conditions}\label{sec:relvolpbc}

\begin{figure*}[t]
\centering
\includegraphics[width=\textwidth]{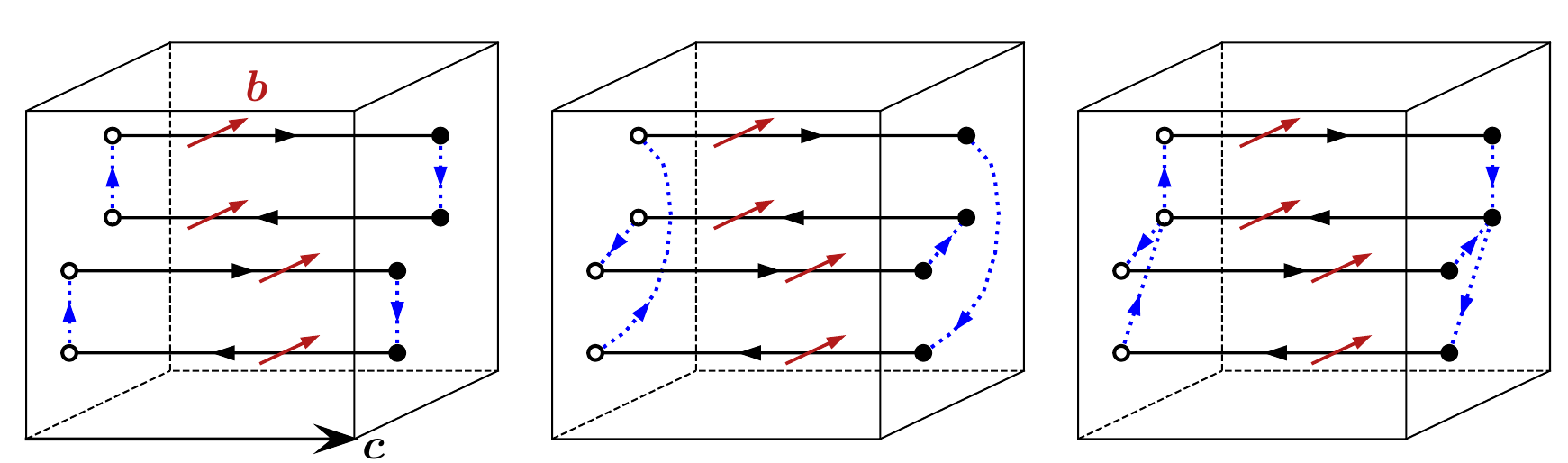}
\caption{Three example closures of an extended dislocation network consisting of two dislocation dipoles in a box with periodic boundary conditions. The network is closed by augmenting the dislocation lines with extra segments, showed by dotted blue lines, that connect the dislocation line ends, with line sense and Burgers vector chosen in accordance with the Burgers vector conservation law at junctions. As the closure lines cancel with themselves upon the repetition of the motif along the periodic cell vector $\vec{c}$, all the examples of closure are equivalent.}
\label{fig:pbc_closure} %% label for entire figure
\end{figure*}

In computational studies of complex dislocation networks, periodic boundary conditions are commonly employed to represent systems of much larger size than the size of the actual simulation cell. While any dislocation network contained in a periodic cell must also be closed by definition, the treatment of periodic boundary conditions requires an additional discussion.

First, networks crossing periodic boundaries must be correctly treated as closed. Depending on the implementation of periodic boundary conditions, line segments crossing a periodic cell boundary may be split and displaced by the corresponding periodic cell vector, seemingly appearing as loose ends. In order to avoid errors, the area vectors must be computed using an unbroken (unwrapped) representation of the network. This is either achieved by computing the distance vector between any two neighbouring points on a curve modulo the periodic cell vectors \cite{bulatov2000periodic}, or by explicitly translating the segments broken by a periodic boundary by the corresponding cell vector.

Second, the simulation box may contain networks that span the entire periodic box along one or more directions, regardless of how they are translated. In an unwrapped representation, such a network appears to contain terminated lines that only become connected if the network is repeated along the periodic cell directions. The simplest example of such a network is a pair of dislocation lines, collinear to one of the periodic cell translation vectors, with identical Burgers vectors and opposing line sense. In what follows, we refer to such a network as \textit{extended}. Extended networks have well-defined relaxation volumes, provided that we connect the loose segments in such a way that the network can be represented by a superposition of closed loops. 

Specifically how should an extended network be closed? As it turns out, a specific recipe for doing this is immaterial, provided that some simple rules are followed. Consider a network extended along a periodic translation vector $\vec{c}$, containing a set of $m$ curved segments terminating at points $\vec{R}_i$ that are connected through translations by $\pm\vec{c}$. We introduce a new set of dislocation lines, the \textit{closure}, linking the loose ends of line segments in a manner such that the Burgers vector conservation law is fulfilled at every junction. Some examples of closure are shown in Fig.~\ref{fig:pbc_closure}. The vector area of the closed network is invariant with respect to the choice of closure since each closure segment cancels itself out with its own periodic image. The most trivial choice of closure is to pick a single loose end of a dislocation segment and let all other ends of loose segments connect to it, keeping the line sense and Burgers vector unchanged, see Fig.~\ref{fig:pbc_closure} (right). The contribution of the closure to the relaxation volume is then given by:
\begin{equation}\label{eq:pbc_network_integral}
\Omega^\mathrm{(c)} = \frac{1}{2}\sum_{i=2}^m s_i \vec{b}^{(i)}\cdot \left( \int_{(i,1)} \vec{r}\times \dint{\vec{r}} + \int_{(1,i)} (\vec{r}-\vec{c})\times \dint{\vec{r}}  \right),
\end{equation}
where integrals over $(i,j)$ signify curve integral starting at $\vec{R}_i$ and ending at $\vec{R}_j$, and $s_i$ is either $-1$ or $1$ depending on the line sense of the loose segment terminating at $\vec{R}_i$. After some manipulation, we arrive at a simple expression
\begin{equation}
\Omega^\mathrm{(c)} = \frac{1}{2}\sum_{i=2}^m s_i \vec{b}^{(i)}\cdot \left[\vec{c} \times \left(\vec{R}_i - \vec{R}_1\right)\right]. \label{eq:correction}
\end{equation}
This closure correction needs to be included in the formula for the relaxation volume given by equation \eqref{eq:volume_network_integral} if a dislocation configuration happens to contain loose ends because of periodicity, to ensure consistency with the treatment of a fully closed dislocation network. The relaxation volume and dipole tensors are then computed in exactly the same way as for a closed network. In practice, we found the term given by (\ref{eq:correction}) to be of comparable magnitude to the relaxation volume of the network itself, and as such it cannot be neglected.

\section{Applications}\label{sec:applications}

\subsection{The relaxation volume of a stacking fault tetrahedron}

Stacking fault tetrahedra (SFT), first observed in quenched gold by Silcox and Hirsch \cite{Silcox1959}, are thought to form through the condensation of individual vacancies into a platelet that subsequently collapses into a loop of intrinsic fault \cite{Kiritani1997,HullBacon2001}. Possible alternative reaction pathways, leading to the formation of SFTs and likely dominating the dynamics of defects in the mesoscopic limit, were extensively analyzed and reviewed in Ref.~\cite{Uberuaga2007}. The faulted loop is able to transform into an SFT by means of dissociation and glide of the dislocation segments bounding the loop. As these processes do not involve mass transfer, in the limit where dislocations are treated as elastic discontinuities and core effects are neglected, the relaxation volume of the SFT configuration is not expected to change during the transformation. The individual components of the dipole and relaxation volume tensors may however change. 

\begin{figure*}[t]
\includegraphics[width=\textwidth]{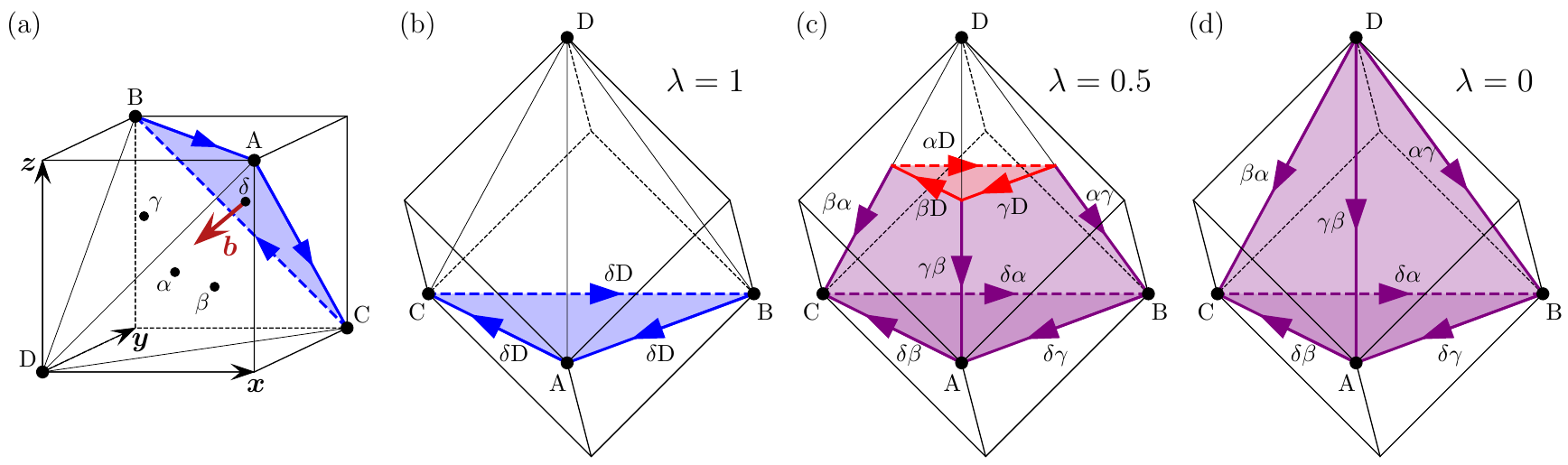}
\caption{(a) Arrangement of the faulted Frank loop and stacking fault tetrahedron corner points in the one-eighth of an fcc unit cell with the edge length of $a/2$. Points labeled by Greek letters are at midpoints of tetrahedral faces opposite to points labeled by the respective Roman letters. (b-d) Transformation of the faulted Frank loop to a stacking fault tetrahedron parameterised by the reaction coordinate $\lambda$. Burgers vectors are described using the Thompson vector notation.}
\label{fig:sft} %% label for entire figure
\end{figure*}

Let the faulted Frank loop be of vacancy type with $\vec{b} = \frac{1}{3}[\overline{111}]$. The loop lies in the plane perpendicular to the Burgers vector, bounded by dislocations with $\left\langle 110\right\rangle$ line directions and $\{112\}$ slip planes forming an equilateral triangle with side length $l$, see Figure~\ref{fig:sft}. We introduce a reaction coordinate $\lambda$, parameterising the transformation of the faulted loop at $\lambda=1$ to an SFT at $\lambda=0$, with $1 > \lambda > 0$ describing the intermediate stages of the transformation process. Intermediate configurations are approximated by a tetrahedron truncated at a height $h'=(1-\lambda)h$ along the direction of $\vec{b}$, see Figure~\ref{fig:sft} illustrating the notations. Here, $h$ is the height of a perfect SFT.

Using the parameterisation of the SFT transformation detailed in Appendix \ref{sec:appendix}, we find the relaxation volume tensor from expression~\eqref{eq:reltensorline} 
\begin{equation}
\tensor{\Omega}(\lambda) = -\frac{a l^2}{12}
\begin{bmatrix}
    1           & \lambda^2  & \lambda^2 \\
    \lambda^2   & 1          & \lambda^2 \\
    \lambda^2   & \lambda^2  & 1
\end{bmatrix},
\end{equation}
with the relaxation volume $\Omega(\lambda) = \Omega_{ii} = -a l^2/4$, where $a$ is the fcc lattice constant. The principal values of the relaxation volume tensor are 
\begin{equation}
\tensor{\Omega}_\mathrm{p.v.} = -\frac{a l^2}{12} \left(1 - \lambda^2, 1 - \lambda^2, 1 + 2\lambda^2 \right),
\end{equation}
indicating that an SFT ($\lambda =0$) generates isotropic volume contraction, whereas a Frank dislocation loop ($\lambda =1$) gives rise to anisotropic volume contraction. While the principal values of the relaxation volume tensor vary as functions of $\lambda$, the total relaxation volume of the SFT is the same as the relaxation volume of the Frank loop. 

Using the stiffness tensor for a cubic crystal \cite{sutton2020physics}\footnote{No summation over repeated indices in the first term here.}
\begin{equation}
\begin{aligned}
C_{ijkl} &= \left(c_{11} - c_{12} - 2 c_{44}\right) \delta_{ij}\delta_{jk}\delta_{kl} \\
    &\phantom{=}+ c_{12} \delta_{ij}\delta_{kl}
     + c_{44}\left(\delta_{ik}\delta_{jl} + \delta_{il}\delta_{jk}\right),
\end{aligned}
\end{equation}
we arrive at an expression for the elastic dipole tensor of an SFT using either equation \eqref{eq:omegatop} or equation \eqref{eq:dipoleline}, namely
\begin{equation}
\tensor{P}(\lambda) = -\frac{a l^2}{12} 
\begin{bmatrix}
    c_{11} + 2c_{12}    & 2c_{44}\lambda^2      & 2c_{44}\lambda^2  \\
    2c_{44}\lambda^2    & c_{11} + 2c_{12}      & 2c_{44}\lambda^2  \\
    2c_{44}\lambda^2    & 2c_{44}\lambda^2      & c_{11} + 2c_{12}
\end{bmatrix}.
\end{equation}
The number of vacancies $N_\mathrm{v}$ forming the SFT can be found by dividing its relaxation volume by the volume of a missing atom in an fcc lattice $-a^3/4$, resulting in  $N_\mathrm{v} = l^2/a^2$. The relaxation volume of an SFT, expressed in terms of the number of vacancies that it contains, is 
\begin{equation}
    \Omega =-\frac{1}{4}N_\mathrm{v} a^3.\label{eq:volume_SFT_Nv}
\end{equation}
For comparison, the relaxation volume of a spherical void in an fcc metal, containing the same number of vacancies, is \cite{mason2019relaxation}
\begin{equation}\label{eq:volume_void_Nv}
    \Omega _{\rm void} = -\left(\frac{3}{2}\right)^{5/3}\pi^{1/3}
        \left(\frac{1-\nu}{1+\nu} \right)\frac{sa^2}{\mu} N_\mathrm{v}^{2/3},
\end{equation}
where $s$ is the surface stress, $\mu$ is the shear modulus and $\nu$ is the Poisson ratio.
Equations \eqref{eq:volume_SFT_Nv} and \eqref{eq:volume_void_Nv} suggest that in the macroscopic limit $N_\mathrm{v} \gg 1$, an SFT exhibits a much greater degree of elastic relaxation than a void, in agreement with atomistic simulations \cite{Uberuaga2007}. This does not imply, however, that a void is energetically more stable than an SFT since, like for point defects \cite{Han2002,NguyenManh2006}, it is often the core energy rather than the elastic energy that dominates the total self-energy of a defect configuration. 

\subsection{\texorpdfstring{$\mathbf{\langle 111\rangle/\langle 100 \rangle}$}{<111>/<100>} bcc tetrahedron}

\begin{figure*}[t]
\centering
\includegraphics[width=\textwidth]{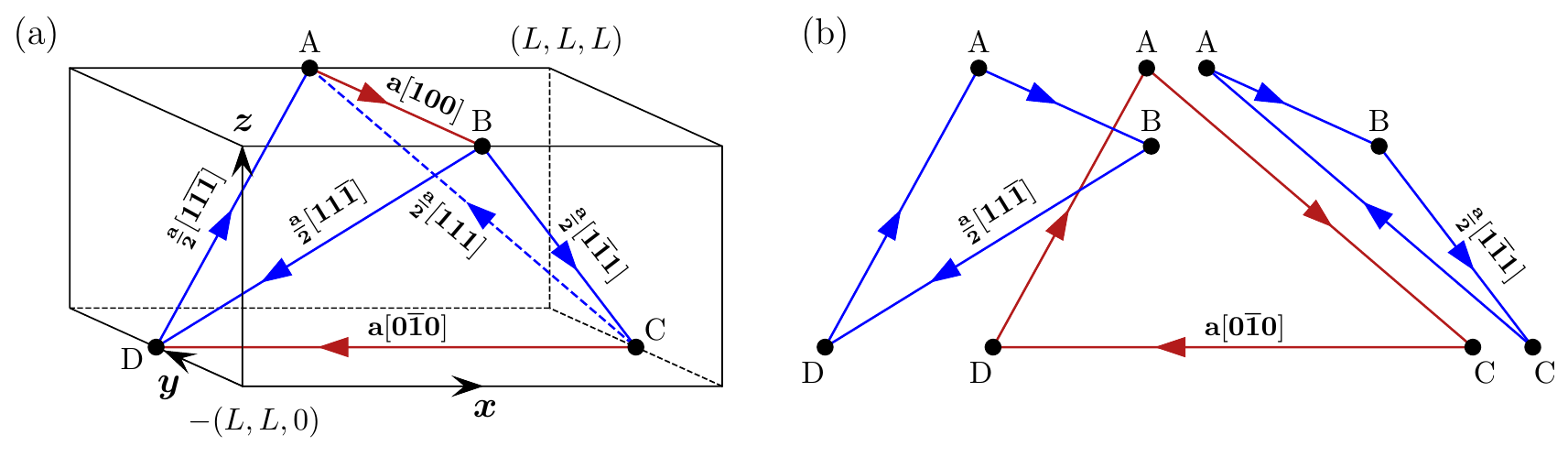}
\caption{(a) Sketch of a tetrahedral defect structure formed by dislocations with $(a/2)\langle 111\rangle$ (blue) and $a\langle 100\rangle$ (red) Burgers vectors in a bcc metal. The configuration resembles a stacking fault tetrahedron, but involves no stacking faults. The topology of this dislocation configuration is not equivalent to that of a dislocation loop. Coordinates of points A, B, C and D are given in Table \ref{tab:bcc_tetra}. (b) Dislocation configuration shown in (a) represented as an equivalent superposition of two $(a/2)\langle 111\rangle$ and one $a\langle 100\rangle$ loop.}
\label{fig:bcc_tetrahedron} %% label for entire figure
\end{figure*}

Collision cascades in body-centered cubic (bcc) metals occasionally produce unusually complex dislocation structures, which are topologically different from the commonly occurring individual dislocation loops. An example of such a topologically unusual dislocation configuration formed in a collision cascade is given in Figure~10 of Ref.~\cite{Sand2018}. Similarly to an SFT, the structure involves four dislocation junctions but, as opposed to an SFT, now dislocations with two fundamentally different Burgers vector types are involved in the formation of this structure. These two Burgers vectors, schematically noted in Figure~\ref{fig:bcc_tetrahedron}, are of the $(a/2)\langle 111 \rangle $ and $a\langle 100 \rangle $ types, where $a$ is the bcc lattice constant. 

Computing the scalar triple products for the six segments of the bcc tetrahedron using the rule \eqref{eq:relaxline} produces values given in the last column of Table \ref{tab:bcc_tetra}, resulting in the total relaxation volume of the dislocation structure in Figure~\ref{fig:bcc_tetrahedron}
\begin{equation}
    \Omega=aL^2.\label{eq:volume_bcc_tetrahedron}
\end{equation}
The fact that this quantity is positive suggests that the dislocation configuration depicted in Figure~\ref{fig:bcc_tetrahedron} has the net interstitial, rather than vacancy, character.

The decomposition of the structure shown in Figure~\ref{fig:bcc_tetrahedron} into individual loops highlights the subtlety associated with the evaluation of the volume of a complex dislocation configuration. Indeed, applying vector calculus and formula $({\bf b}\cdot {\bf A})$ for the volume of an individual loop \eqref{eq:volume_loop}, we find that both $(a/2)\langle111\rangle$ dislocation loops, shown in blue, have the interstitial character and the same positive volumes of $aL^2$. The $a\langle100\rangle$ loop, shown in red, has vacancy character, its volume is negative, and is equal to $-aL^2$. The sum of volumes of the three loops is $aL^2$, in agreement with the result  \eqref{eq:volume_bcc_tetrahedron} obtained by a direct calculation using equation \eqref{eq:relaxline}, which notably does not require decomposing a dislocation configuration into individual loops.

\subsection{An extended network of dislocations produced by irradiation}

Metals exposed to irradiation by highly energetic particles, such as neutrons in a nuclear reactor, develop microstructures often involving complex extended dislocation network configurations. Whenever energetic particles scatter by atoms in the crystal lattice, they may impart sufficient kinetic energy to ballistically displace a large number of atoms from their lattice sites. The displaced atoms eventually recrystallise, leaving behind defects of both vacancy and interstitial type. With increasing exposure to radiation, the density of radiation defects becomes large enough for defects to coalesce and form complex interconnected dislocation networks. The dislocation network transfers mass through the system under the influence of irradiation and external stress, thereby contributing to irradiation-induced creep.

\begin{figure*}[t]
\includegraphics[width=\textwidth]{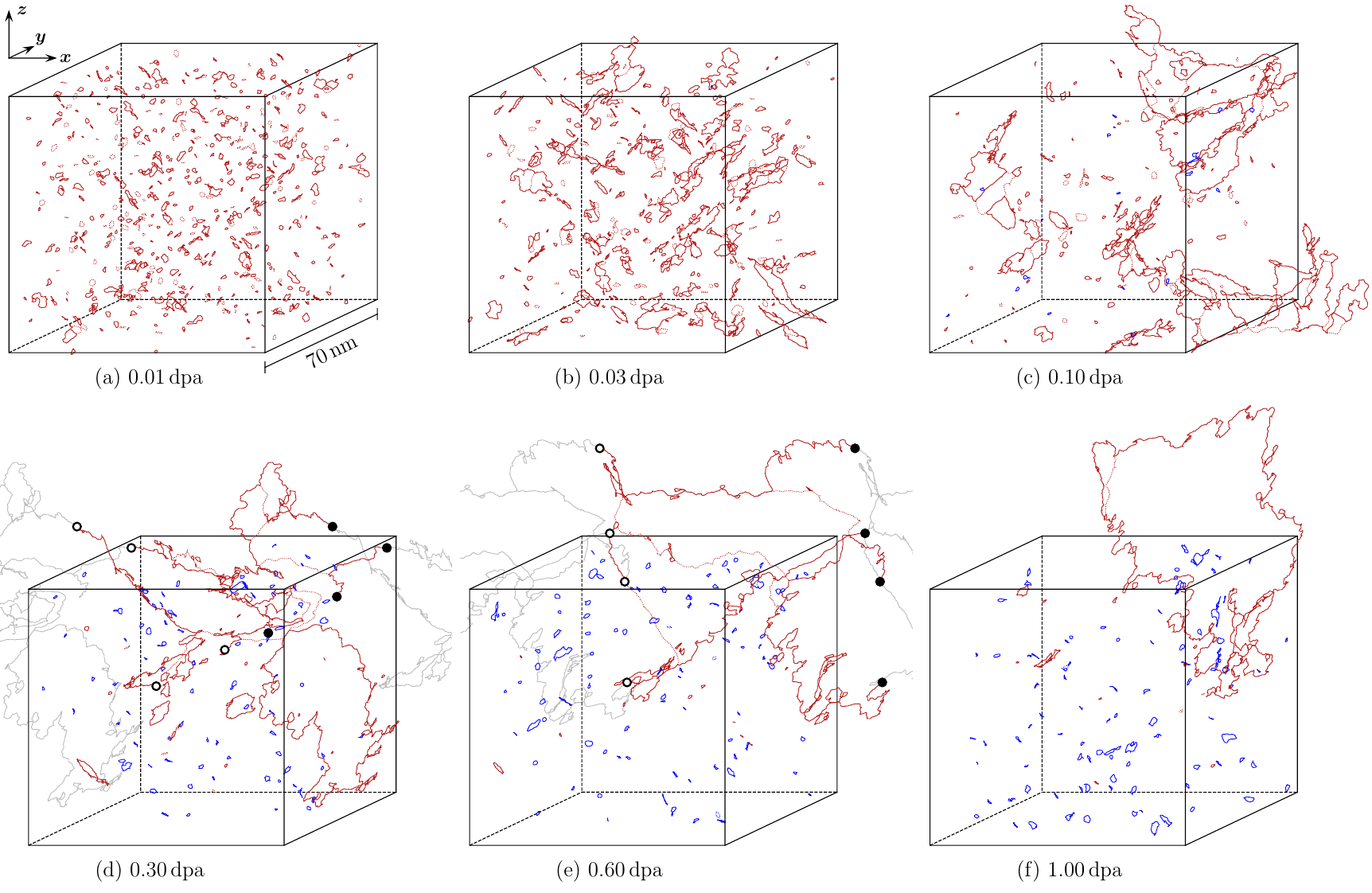}
\caption{Dislocation networks extracted from simulated microstructure of irradiated tungsten, showed in an unwrapped representation. In this example, the dislocation microstructure is found to evolve from dispersed loops at low dose, to extended dislocation networks at intermediate dose, to a large closed interstitial loop at high dose. Networks with net interstitial and vacancy content are colored red and blue, respectively. Structures (d) and (e) contain extended networks, shown periodically repeated along $x$-direction (gray lines), with segment end points marked by black dots. Dislocations of $(a/2)\langle 111 \rangle$ and $a\langle 100 \rangle$ type are drawn with solid and dotted lines, respectively.}
\label{fig:dxadislocations} %% label for entire figure
\end{figure*}

We simulate the formation of an irradiated microstructure in single-crystal bcc tungsten (W) using the molecular dynamics software package \textsc{Lammps} \cite{plimpton1995fast} and the empirical interatomic potential developed in Ref.~\cite{mason2017empirical}. We start by initialising a perfect bcc lattice in an orthogonal simulation cell with 220 unit cell repeats along the $[100]$, $[010]$, and $[001]$ cell vectors, containing over 21.3 million atoms. Periodic boundary conditions  are applied in all three directions. We introduce 100 simultaneous cascades by choosing 100 random atoms, with a minimal distance of \SI{15}{\angstrom} to each other, and assigning each atom a randomly oriented velocity corresponding to the kinetic energy of $\SI{10}{\keV}$. Following the method commonly used in atomistic cascade simulation, we use an electronic stopping model for atoms with kinetic energies above $\SI{10}{eV}$ and an adaptive time step. For atoms with kinetic energy below the energy corresponding to melting, we apply a damping term with the time-constant of $\SI{15.6}{\ps}$ to model electronic losses due to electron-phonon coupling in the low velocity limit \cite{mason2015incorporating}. The system is propagated until the temperature drops below \SI{100}{\K}, which takes $\sim$\SI{10}{\ps}, at which point velocities are set to zero, and atomic coordinates and box dimensions are relaxed using the Conjugate Gradients method. The next round of cascades is initialised in the now damaged microstructure, and this process is repeated until the desired radiation dose is reached. 

Using the damage model proposed in Ref.~\cite{norgett1975proposed}, we find that every new set of cascades increments the radiation dose $\phi$ by the amount
\begin{equation}
\Delta\phi = N_\mathrm{c}\frac{0.8 T_\mathrm{d}}{2 E_\mathrm{d} N} = \SI{1.67e-4}{dpa},
\end{equation}
where $N_\mathrm{c}$ is number of cascades per iteration, $T_\mathrm{d} \approx \SI{8}{\keV}$ is the cascade damage energy, $E_\mathrm{d} = \SI{90}{\eV}$ is the assumed threshold displacement energy for tungsten, and $N$ is the total number of atoms in the simulation cell. The dose is given in the dimensionless units of \textit{displacements per atom} (dpa), which is a standard measure of exposure of materials to radiation used in the field of nuclear materials \cite{norgett1975proposed}. The dose of $\phi = \SI{1}{dpa}$ signifies that on average every atom in the system has been a part of a Frenkel pair, consisting of a vacancy and a self-interstitial defect, suggesting a significant accumulation of radiation defects. We apply 6000 cascade iterations in order to reach the dose of \SI{1}{dpa}.

The resulting atomic configurations are analysed and their dislocation content determined using the Dislocation Extraction Algorithm (DXA) \cite{stukowski2012automated} implemented in the \textsc{Ovito} software \cite{stukowski2009visualization}. The algorithm is able to identify the dislocation line sense, its Burgers vector, the line coordinates, and the topology of the dislocation network. Because the irradiated microstructure is vacancy-rich as a large number of self-interstitial atom defects have coalesced into dislocations loops and the dislocation network \cite{Derlet2020}, some dislocation curves identified by the DXA are disjoint and have loose ends. To ensure that all the networks are closed, we restore the dislocation connectivity by iterating over loose ends in the network and connecting them to other loose ends in the neighbourhood.

\begin{figure*}[t]
\centering
\includegraphics[width=\textwidth]{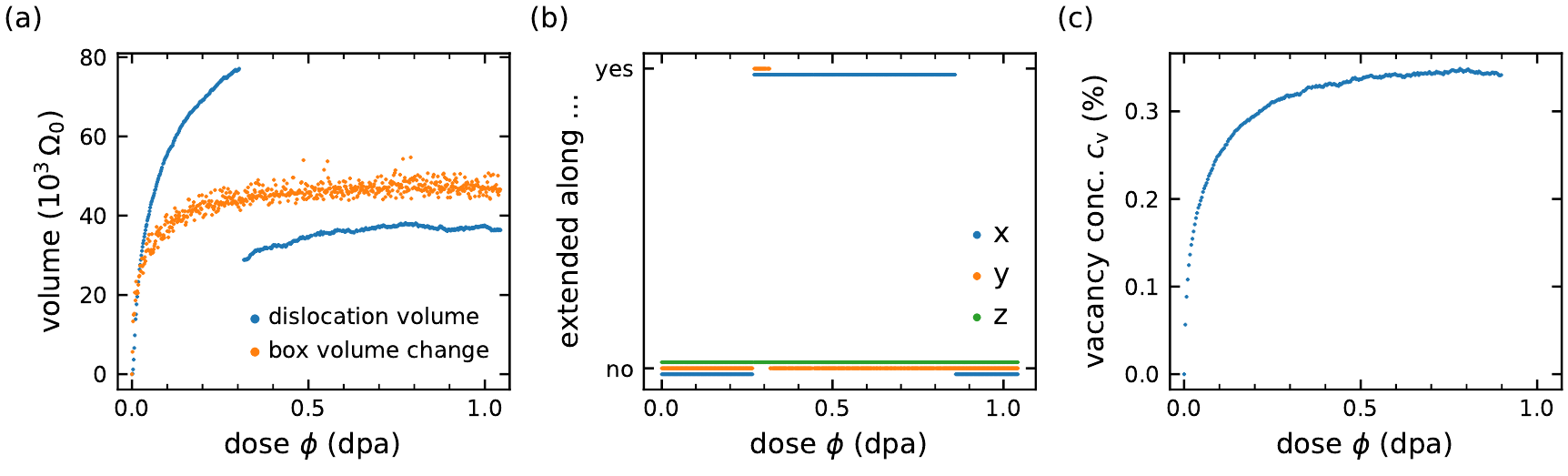}
\caption{(a) The relaxation volume of the dislocation network and the total change of volume of the simulation cell, expressed in the units of atomic volume $\Omega_0$, shown as a function of dose. (b) Plot showing if the dislocation network is extended or not as a function of dose. The discontinuity in the computed relaxation volume occurs shortly after the dislocation network becomes extended, suggesting that the change of the relaxation volume is associated with the change in the topology of the network. (c) Vacancy concentration, in the units of vacancies per lattice site, shown as a function of dose. 
}
\label{fig:dxaresults} %% label for entire figure
\end{figure*}

Snapshots of the dislocation network formed at various doses are shown in Figure~\ref{fig:dxadislocations}. At low doses, the interstitial defects produced by cascades form small spatially dispersed dislocation loops with $(a/2)\langle 111 \rangle$ and $a\langle 100 \rangle$ Burgers vectors. As the dose increases, the interstitial loops grow and coalesce, joining together to form more complex loops, which eventually merge to form an extended network. At a higher dose ($\phi \sim \SI{1}{dpa}$) the network breaks apart, leaving behind a large interstitial loop of around \SI{60}{\nm} diameter. We note that the entirety of the simulation box is now filled with homogeneously distributed vacancy clusters, saturating to the total vacancy content of \SI{0.34}{\percent} over the course of irradiation, see Figure~\ref{fig:dxaresults}c. Around \SI{90}{\percent} of vacancy content is in the form of mono-vacancies, with the remainder constituting sub-nanometer sized vacancy clusters. The vacancy content was determined using the isosurface method presented in \cite{mason2021parameter}.

In Figure~\ref{fig:dxaresults}a, we compare the change of volume of the simulation cell with the total relaxation volume of the dislocation network computed using expressions~\eqref{eq:volume_network_integral} and \eqref{eq:pbc_network_integral}. The volume change is given in the units of bcc atomic volume $\Omega_0 = a^3/2$. In the dynamic steady state of the material forming in the high dose limit \cite{Derlet2020,Mason2020}, the total box volume has increased by $\Delta V/V_0 = 46\times 10^3/(2\times 220^3) = \SI{0.22}{\percent}$. We note that the relaxation volume of the dislocation network is at times higher than the total volume change of the simulation cell. This is not a contradiction, as each of the many monovacancies contributes the negative amount of $\Omega_\mathrm{mono}^\mathrm{vac} = -0.367\, \Omega_0$ to the total relaxation volume of the system \cite{mason2019relaxation,MaPRM2019a}. For the saturated vacancy concentration of $\SI{0.34}{\percent}$, we obtain the total relaxation volume of all the vacancies $\Omega_\mathrm{tot}^\mathrm{vac} = -26.6\times 10^3\, \Omega_0$, suggesting that the sum of the dislocation and vacancy relaxation volumes is always smaller than the simulation cell volume change. The margin of error stems from neglecting the non-linear relaxation volume effects \cite{mason2017empirical} associated for example with the core regions of dislocations.

The relaxation volume of the dislocation network exhibits a discontinuity at \SI{0.3}{dpa} where it sharply drops by $\Delta\Omega = -48.2\times 10^3\, \Omega_0$. This drop can be entirely attributed to its transformation into an extended network. As seen in Figure~\ref{fig:dxaresults}b, the dislocation network first becomes simultaneously extended along the $x$ and $y$ directions at $\SI{0.26}{dpa}$. Shortly after, at \SI{0.30}{dpa}, the network reorganises itself to being only extended in the $x$ direction. During this process, it loses over half of its relaxation volume. Because the number of atoms in the system is conserved, such a process can only occur if the crystal lattice simultaneously undergoes a plastic transformation that increases the number of lattice sites by an equal amount --- this is possible because many lattice sites in the simulation cell are unoccupied, populated by vacancies. As the simulation cell volume does not exhibit a discontinuity, the drop in the relaxation volume of the dislocation network must be fully counteracted by the volume change arising from a corresponding plastic deformation of the cell as a whole.

To illustrate this point, we estimate the change in the number of lattice sites in the simulation cell. Given matrix $\tensor{A}$, where the $j$-th column is the $j$-th simulation cell vector, and matrix $\tensor{B}$, where the $j$-th column is the $j$-th primitive lattice vector, preserving the continuity across periodic boundary conditions requires that
\begin{equation}\label{eq:continuity1}
    \tensor{A} = \tensor{B}\tensor{N},
\end{equation}
where $\tensor{N}$ is a 3$\times$3 matrix of integers. Note that $n_\mathrm{lattice} = n_\mathrm{motif}\det(\tensor{N}) $ is the number of lattice sites, where $n_\mathrm{motif} = 1$ for the primitive bcc unit cell. For a bcc crystal, $\tensor{B}$ takes on the form
\begin{equation}
    \tensor{B} = \frac{a}{2}
            \left[ \begin{array}{ccc}
                 -1            & \phantom{-}1  & \phantom{-}1     \\
                 \phantom{-}1  & -1            & \phantom{-}1    \\
                 \phantom{-}1  & \phantom{-}1  & -1
            \end{array} \right].
\end{equation}
For the simulation cell given here before irradiation, $A_{ij} = 220 a \delta_{ij}$, and therefore $N_{ij} = 220 (1 - \delta_{ij})$.

Consider now the same system after it was exposed to irradiation, leading to an elasto-plastic deformation, transforming the simulation cell to $\tensor{A}'$. As the periodic boundary conditions still apply to the transformed simulation cell, the continuity condition still applies in a modified form
\begin{equation}\label{eq:continuity2}
    \tensor{A}' = \tensor{B}'\tensor{N}',
\end{equation}
where $\tensor{B}'$ is the elastically distorted lattice cell and $\tensor{N}'$ is the integer-valued matrix of cell repeats after plastic deformation. The change in the number of lattice sites after elasto-plastic deformation is then given by 
\begin{equation}
    \Delta n_\mathrm{lattice} = n_\mathrm{motif} \left[\det{(\tensor{N}')} - \det{(\tensor{N})}\right].
\end{equation}
As the continuity condition still applies, $\tensor{N}'$ is constrained to be integer-valued, therefore the smallest possible increase in the number of lattice sites is $\Delta n_\mathrm{lattice} = 48400$, which is equivalent to the volume increase of the system by $\Delta V = 48.4\times 10^3\, \Omega_0$. This volume change is very close to the magnitude of the discontinuous change in the dislocation network volume of $\Delta \Omega = -48.2 \times 10^3\, \Omega_0$ found numerically at $\phi=\SI{0.3}{dpa}$. This confirms that the change of volume of the crystal involves a plastic deformation, mediated by a structural reorganisation of the extended dislocation network. During irradiation at low temperatures, where vacancies are immobile, this mechanism is expected to play a significant role in changing the dimensions of the crystal.

\section{Conclusions}

The question about the relaxation volume of an arbitrarily complex interconnected network of dislocations naturally arises in the context of macroscopic analysis of microstructures of heavily irradiated or heavily deformed materials if one attempts to evaluate the effect of the dislocation network on swelling or dimensional changes \cite{dudarev2018multi,reali2021macroscopic}. While a formula for the volume of an isolated dislocation loop is well known \cite{Dederichs1973,dudarev2018elastic}, generalizing it to the case of an arbitrary dislocation network has so far proved elusive. In this study, we derived an analytical expression for the volume of an arbitrary interconnected network of dislocations and showed that its volume can be evaluated using piece-wise line integration along the arbitrarily curved directionally ordered dislocation lines, linking the junctions of the network, see Eqns. \eqref{eq:relaxline} and \eqref{eq:volume_network_integral}. We prove that this analytical expression for the volume of the network is invariant with respect to the choice of the Cartesian system of coordinates or the use of periodic boundary conditions, and illustrate its applications using several representative examples of complex dislocation structures.

\begin{acknowledgments}

We are grateful to J. Marian, S. J. Zinkle, A. E. Sand, B. Uberuaga, A. Stukowski and A. P. Sutton for stimulating discussions. %
This work has been carried out within the framework of the EUROfusion Consortium, funded by the European Union via the Euratom Research and Training Programme (Grant Agreement No 101052200 - EUROfusion), and by the RCUK Energy Programme, grant No. EP/W006839/1. Enrique Mart\'inez acknowledges support from the startup package at Clemson University, and Sergei Dudarev acknowledges support from the Centre for Non-Linear Studies at LANL.
Views and opinions expressed are however those of the authors only and do not necessarily reflect those of the European Union or the European Commission. Neither the European Union nor the European Commission can be held responsible for them. %
The authors acknowledge the use of the Cambridge Service for Data Driven Discovery (CSD3) and associated support services provided by the University of Cambridge Research Computing Services (\href{www.csd3.cam.ac.uk}{www.csd3.cam.ac.uk}) that assisted the completion of this study.

%This work was performed using resources provided by the Cambridge Service for Data Driven Discovery (CSD3) operated by the University of Cambridge Research Computing Service (\href{www.csd3.cam.ac.uk}{www.csd3.cam.ac.uk}), provided by Dell EMC and Intel using Tier-2 funding from the Engineering and Physical Sciences Research Council (capital grant EP/T022159/1), and DiRAC funding from the Science and Technology Facilities Council (\href{www.dirac.ac.uk}{www.dirac.ac.uk}).

\end{acknowledgments}

\appendix

\section{Parameterisation of the SFT formation process}\label{sec:appendix}

Following Fig.~\ref{fig:sft}, the corners of the one-eighth fcc unit cell $\vec{A}$, $\vec{B}$, $\vec{C}$, and $\vec{D}$ define the coordinate system:
\begin{equation}
\begin{aligned}
\vec{A} &=  \frac{a}{2}\left(\vec{\hat{x}} + \vec{\hat{z}}\right) \\
\vec{B} &=  \frac{a}{2}\left(\vec{\hat{y}} + \vec{\hat{z}}\right) \\
\vec{C} &=  \frac{a}{2}\left(\vec{\hat{x}} + \vec{\hat{y}}\right) \\
\vec{D} &=  0,
\end{aligned}
\end{equation}
% \frac{l}{\sqrt{2}} 
%
where $a$ is the fcc lattice constant. Points $\bm{\alpha}$, $\bm{\beta}$, $\bm{\gamma}$, and $\bm{\delta}$ lie at midpoints of the tetrahedral faces opposite the points $\vec{A}$, $\vec{B}$, $\vec{C}$ and $\vec{D}$, respectively:
\begin{equation}
\begin{aligned}
\bm{\alpha} &=  \left(\vec{B} + \vec{C} + \vec{D}\right)/3 \\
\bm{\beta}  &=  \left(\vec{C} + \vec{D} + \vec{A}\right)/3 \\
\bm{\gamma} &=  \left(\vec{D} + \vec{A} + \vec{B}\right)/3 \\
\bm{\delta} &=  \left(\vec{A} + \vec{B} + \vec{C}\right)/3 .
\end{aligned}
\end{equation}
{Tables~\ref{tab:parafrank}-\ref{tab:parasft}} list the Burgers vectors $\vec{b}^{(n)}$, starting points  $\vec{p}^{(n)}$, and end points $\vec{q}^{(n)}$ of the piecewise linear segments constituting the SFT dislocation structure.  Note that the Burgers vectors are given in the Thompson vector notation, such that
\begin{equation}
\vec{b} = \bm{\delta \mathrm{D}} = \vec{D} - \bm{\delta} = \frac{1}{3} [\overline{111}].
\end{equation}
The Tables enable a simple computation of the dipole tensor, the relaxation volume tensor, and the relaxation volume, following the general expressions {\eqref{eq:dipoleline}-\eqref{eq:relaxline}}.

\begin{table}[t]
  \caption{Parameterisation of the faulted Frank loop ($\lambda = 1$) in terms of piecewise linear segments.\label{tab:parafrank}}
\begin{ruledtabular}  
\begin{tabular}{ccc}
  	Burgers vector $\vec{b}^{(n)}$ ($a$)    
  	        & start $\vec{p}^{(n)}$ ($\sqrt{2}l$) & end $\vec{q}^{(n)}$ ($\sqrt{2}l$)\\
   \hline
  	$\bm{\delta\mathrm{D}} = \frac{1}{3}[\overline{111}]$                     
  	        &  $\vec{A}$          &  $\vec{C}$        \\
    $\bm{\delta\mathrm{D}} = \frac{1}{3}[\overline{111}]$                     
            &  $\vec{C}$          &  $\vec{B}$        \\
    $\bm{\delta\mathrm{D}} = \frac{1}{3}[\overline{111}]$                     
            &  $\vec{B}$          &  $\vec{A}$        \\
  	\end{tabular}
\end{ruledtabular}
\end{table}

\begin{table}[t]
\caption{Parameterisation of the Frank loop transforming to the SFT ($0 \leq \lambda < 1$) in terms of piecewise linear segments.\label{tab:parasft}}
\begin{ruledtabular}  
\begin{tabular}{ccc}
  	Burgers vector $\vec{b}^{(n)}$ ($a$)    
  	        & start $\vec{p}^{(n)}$ ($\sqrt{2}l$) & end $\vec{q}^{(n)}$ ($\sqrt{2}l$)\\
   \hline
	$\bm{\delta\beta}  = \frac{1}{6}[0\overline{1}\overline{1}]$                     
  	        &  $\vec{A}$          &  $\vec{C}$        \\
    $\bm{\delta\alpha} = \frac{1}{6}[\overline{1}0\overline{1}]$                      
            &  $\vec{C}$          &  $\vec{B}$        \\
    $\bm{\delta\gamma} = \frac{1}{6}[\overline{1}\overline{1}0]$                      
            &  $\vec{B}$          &  $\vec{A}$        \\
    $\bm{\gamma\beta} = \frac{1}{6}[10\overline{1}]$                      
            &  $ \vec{D} + \lambda(\vec{A}-\vec{D})$          &  $\vec{A}$        \\
    $\bm{\beta\alpha} = \frac{1}{6}[\overline{1}10]$                      
            &  $ \vec{D} + \lambda(\vec{C}-\vec{D})$          &  $\vec{C}$
                \\
    $\bm{\alpha\gamma} = \frac{1}{6}[0\overline{1}1]$                      
            &  $ \vec{D} + \lambda(\vec{B}-\vec{D})$          &  $\vec{B}$        \\
	$\bm{\beta\mathrm{D}}  = \frac{1}{6}[\overline{211}]$                     
  	        &  $\vec{D} + \lambda(\vec{A}-\vec{D})$          
  	            & $\vec{D} + \lambda(\vec{C}-\vec{D})$ \\
    $\bm{\alpha\mathrm{D}} = \frac{1}{6}[\overline{121}]$                      
  	        &  $\vec{D} + \lambda(\vec{C}-\vec{D})$          
  	            & $\vec{D} + \lambda(\vec{B}-\vec{D})$ \\
    $\bm{\gamma\mathrm{D}} = \frac{1}{6}[\overline{112 }]$                      
  	        &  $\vec{D} + \lambda(\vec{B}-\vec{D})$          
  	            & $\vec{D} + \lambda(\vec{A}-\vec{D})$ \\
  	\end{tabular}
\end{ruledtabular}
\end{table}

\section{Parameterisation of a bcc tetrahedron}\label{sec:appendix2}

Fig.~\ref{fig:bcc_tetrahedron} shows the schematic structure of a tetrahedral dislocation structure that can be formed in a bcc metal by the ${\bf b}=(a/2)\langle 111 \rangle$ and ${\bf b}=a\langle 001\rangle$ dislocations. In principle, the dislocation lines forming the edges of a bcc tetrahedron can be curved, see for example Ref.~\cite{Sand2018}, but for the purpose of illustrating the principle of how to evaluate the volume of an unusual dislocation structure, we assume that all the dislocation lines linking the dislocation junctions of the structure are straight. The coordinates of the four junctions shown in Fig.~\ref{fig:bcc_tetrahedron} are  
\begin{equation}
\begin{aligned}
\vec{A} &=  (0,L,L) \\
\vec{B} &=  (0,-L,L) \\
\vec{C} &= (L,0,0) \\
\vec{D} &= (-L,0,0).
\end{aligned} \label{eq:bcc_tetrahedron_vertices}
\end{equation}
In these notations, vector $\vec{DC}$ is collinear with the $[100]$ crystallographic direction, whereas vector $\vec{BA}$ is collinear with the $[001]$ direction. The Burgers vectors of the six dislocation segments forming the bcc tetrahedron shown in Fig.~\ref{fig:bcc_tetrahedron}, and the vector coordinates of the segments themselves are given in Table~\ref{tab:bcc_tetra}. A direct examination of the dislocation configuration shows that the Burgers vectors are conserved at junctions, and that the $a\langle 001\rangle$ dislocations have a pure edge character whereas the $(a/2)\langle 111\rangle$ dislocations have a mixed character. The length of the $a\langle 001\rangle$ segments is $2L$ whereas the length of the $(a/2)\langle 111\rangle$ segments is $\sqrt{3}L$.

\begin{table}[t]
\caption{Parameterisation of the bcc tetrahedron shown in Fig.~\ref{fig:bcc_tetrahedron} in terms of piecewise linear dislocation segments.\label{tab:bcc_tetra}}
\begin{ruledtabular}  
\begin{tabular}{cccc}
  	Burgers vector $\vec{b}^{(n)}$ ($a$)     
  	        & start  $\vec{p}^{(n)}$ & end $\vec{q}^{(n)}$ & $\Omega ^{(n)}$ \\
   \hline
  	$[100]$                     
  	        &  $\vec{A}$          &  $\vec{B}$ & $aL^2$        \\
  	$\frac{1}{2}[1{\overline 1}{\overline 1}]$                     
  	        &  $\vec{D}$          &  $\vec{A}$ & $0$       \\
   	$\frac{1}{2}[111]$                     
  	        &  $\vec{C}$          &  $\vec{A}$ & $0$       \\  
    $[0 \overline{1}0]$                     
  	        &  $\vec{C}$          &  $\vec{D}$ & $0$       \\ 
  	$\frac{1}{2}[1 \overline{1}1]$                     
  	        &  $\vec{B}$          &  $\vec{C}$ & $0$   \\ 	
   	$\frac{1}{2}[1 1\overline {1}]$                     
  	        &  $\vec{B}$          &  $\vec{D}$ & $0$ \\  	
  	\end{tabular}
\end{ruledtabular}
\end{table}

The quantities referred to as $\Omega ^{(n)}$ in Table~\ref{tab:bcc_tetra} are the individual terms in equations \eqref{eq:relaxline} and \eqref{eq:volume_network_integral}, computed in the Cartesian system of coordinates where the positions of the junctions are given by \eqref{eq:bcc_tetrahedron_vertices}. On their own, quantities $\Omega ^{(n)}$ are not invariant with respect to the translations of the system of coordinates, but their sum $\Omega = \sum _n \Omega ^{(n)}$ is invariant with the respect to such translations.

\bibliography{main}

\end{document}